\documentclass{amsproc}

\newtheorem{theorem}{Theorem}[section]

\theoremstyle{definition}
\newtheorem{definition}[theorem]{Definition}
\newtheorem{example}[theorem]{Example}

\theoremstyle{remark}
\newtheorem{remark}[theorem]{Remark}

\numberwithin{equation}{section}

\usepackage{amssymb}
\usepackage{mathrsfs}
\usepackage{tikz}
\usetikzlibrary{cd}
\usepackage{xcolor}
\usepackage{subcaption}
\usepackage{caption}

\begin{document}

\title{Link Homology from Homological Mirror Symmetry}

\author{Elise LePage}
\address{Department of Physics, University of California, Berkeley, CA, 94720}
\email{elepage@berkeley.edu}

\date{March 22, 2023 and, in revised form, June 15, 2023.}


\begin{abstract}
	We explain how to calculate link homology for a Lie algebra $\mathfrak{g}$ using the Fukaya category associated to a 2d A-model. Links are represented as configurations of particular A-branes and link homology is given by Homs between these A-branes. In the case of $\mathfrak{g}=\mathfrak{su}_2$, we explain how to explicitly construct projective resolutions of the relevant A-branes in terms of thimbles, whose algebra is known. This gives an explicit algorithm for computing Khovanov homology. This algorithm can be extended to all Lie algebras.
\end{abstract}

\maketitle

\section{Quantum group invariants and categorification}
Given a Lie algebra and a representation, one may define a certain polynomial link invariant, known as a quantum group invariant. The first example is the Jones polynomial, associated to the fundamental representation of $\mathfrak{su}_2$ and defined by Jones in 1984 \cite{Jones}. Four years later, Witten gave a physical origin of the Jones polynomial using Chern-Simons theory for $\mathfrak{su}_2$, which allowed one to define a link invariant for any Lie algebra $\mathfrak{g}$ using Chern-Simons theory for $\mathfrak{g}$ \cite{Witten}. 

A natural question to ask is whether one can categorify quantum group invariants to get new invariants of links. In 1999, Khovanov proposed a categorification for the case of $\mathfrak{su}_2$, now known as Khovanov homology \cite{Khovanov}. Khovanov homology categorifies the Jones polynomial in the sense that it associates to a link a series of homology groups, which are themselves link invariants, whose Euler characteristic coincides with the Jones polynomial of the link. Khovanov and Rozansky extended this categorification to $\mathfrak{su}_n$, defining Khovanov-Rozansky homology \cite{KR}. In 2013, Webster proposed a categorification that works for all Lie algebras \cite{Webster}. Even more recently, Aganagic proposed a solution based in physics using homological mirror symmetry \cite{Aganagic1}, \cite{Aganagic2}. Her work gives the same categorification as Webster's work but has the advantage that the physical origin is clearer and calculations become very tractable. Furthermore, Aganagic's approach extends to Lie superalgebras. Although Aganagic defines link homology in a pair of mirror theories, here we focus only on the A-side of mirror symmetry.  

\section{The A-model}
Consider a simple Lie algebra $\mathfrak{g}$. We want to define a Landau-Ginzburg theory associated to $\mathfrak{g}$. Towards this end, let $\mathcal{A} = \mathbb{C}^\times$ and $Y= \bigotimes_{a=1}^{rk\mathfrak{g}} Sym^{d_a} \mathcal{A}$ where the $d_a$ may take any integer value. We will see shortly that we will require special choices of $d_a$ in order to get link invariants. It is helpful to think of $\mathcal{A}$ as an infinite cylinder and $Y$ as the configuration space of colored points (with $d_a$ points of each color) on the cylinder.

Let $a_i$ be a series of marked points on $\mathcal{A}$, each colored by a representation of $\mathfrak{g}$ with highest weight $\mu_i$. Denote the roots of $\mathfrak{g}$ by $e_a$. Let $y_{\alpha,a}$ be the coordinate on each copy of $\mathcal{A}$ in $Y$, labelled such that $(y_{1,a},y_{2,a},\ldots,y_{d_a,a})$ are coordinates on $Sym^{d_a} \mathcal{A}$. Define a holomorphic function $W:Y \to \mathbb{C}$ by
\begin{equation}
	W(y) = \lambda_0 W^0(y) + \sum_{a=1}^{rk\mathfrak{g}} \lambda_a W^a(y)
\end{equation}
where $\lambda_0, \lambda_a \in \mathbb{C}$ are parameters and
\begin{equation}
	W^0(y) = \sum_{a=1}^{rk\mathfrak{g}} \ln f_a(y), \qquad W^a(y) = \sum_{\alpha=1}^{d_a} \ln y_{\alpha,a}
\end{equation}
with
\begin{equation}
	f_a(y) = \prod_{\alpha=1}^{d_a} \frac{\prod_i (1-a_i/y_{\alpha,a})^{\langle e_a, \mu_i \rangle}}{\prod_{(b,\beta)\neq(a,\alpha)} (1-y_{\beta,b}/y_{\alpha,a})^{\langle e_a, e_b \rangle/2}}.
\end{equation}
We will also need to equip $Y$ with a choice of a top holomorphic form given by
\begin{equation}
	\Omega = \bigwedge_{a=1}^{rk\mathfrak{g}} \bigwedge_{\alpha=1}^{d_a} \frac{dy_{\alpha,a}}{y_{\alpha,a}}.
\end{equation}

We will consider the derived Fukaya-Seidel category $\mathscr{D}_Y$ associated to $(Y,W)$. The objects of this category are A-branes supported on graded Lagrangian submanifolds of $Y$. 

\subsection{The gradings}
Lagrangians on $Y$ admit multiple gradings: a Maslov grading $M$ depending on $\Omega$ and equivariant gradings $J^a$ depending on $W^a$. An A-brane requires both a choice of a Lagrangian and a choice of gradings.

The Maslov grading comes from a choice of phase of $\Omega$ along a Lagrangian $L$. Let $\Omega^2 = |\Omega^2| e^{2i\varphi}$. Let $\tilde{Y}$ be a cover of $Y$ on which $2\varphi$ is single-valued. Let $\tilde{L}$ be a lift of $L$ to $\tilde{Y}$. Let $\tilde{L}[M]$ denote a Lagrangian on which the value of $\varphi$ differs from its value on $\tilde{L}$ by $\pi i M$:
\begin{equation}
	\varphi \vert_{\tilde{L}[M]} = \varphi \vert_{\tilde{L}} + \pi i M.
\end{equation}

The equivariant gradings arise from the non-single-valuedness of $W^a$. More precisely, a shift in the equivariant gradings of $\tilde{L}$ changes the value of $W$ on $\tilde{L}$ as follows:
\begin{equation}
	W^a \vert_{\tilde{L}\{\vec{J}\}} = W^a \vert_{\tilde{L}} + 2 \pi i J^a.
\end{equation}

More details regarding the gradings can be found in \cite{Aganagic2}.
\subsection{Morphisms between A-branes}
Morphisms between A-branes are defined via Floer theory. Let $L_0$ and $L_1$ be two A-branes. Then morphisms from $L_0$ to $L_1$ are given by
\begin{equation}
	Hom_{\mathscr{D}_Y} (L_0, L_1) = HF^{0,0} (L_0, L_1)
\end{equation}
where $HF^{0,0}$ denotes the Floer cohomology groups in Maslov and equivariant degrees zero. See \cite{Auroux} or \cite{OS} for a more detailed review of Floer theory. The groups $HF^{M,\vec{J}} (L_0, L_1)$ are generated by the intersection points of $\tilde{L}_0$ with $\tilde{L}_1[M]\{\vec{J}\}$ in $\tilde{Y}$ modulo the action of a differential $Q$. Equivalently, the intersection points are the degree $(M,\vec{J})$ intersection points of $L_0$ with $L_1$ in $Y$.

\begin{figure}
	\includegraphics{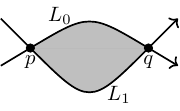}
	\caption{An example of a holomorphic disk contributing to $Qp$}
	\label{fig:disk}
\end{figure}

The Floer differential $Q$ counts certain degree $(1,\vec{0})$ holomorphic disks between intersection points. Let $p, q \in L_0 \cap L_1$. Let $y:[0,1]\times\mathbb{R} \to Y$ be a holomorphic map such that $y(0,t)=L_0$, $y(1,t)=L_1$, $\lim_{t \to \infty} (s,t) = p$, and $\lim_{t \to -\infty} (s,t) = q$. We require that $y^*W$ is a regular function. An example of the image in $Y$ of one such disk is shown in Figure \ref{fig:disk}. Note that by the Riemann mapping theorem, we can just as well consider maps from the unit disk $D$ to $Y$, which we will also call $y$. The Maslov and equivariant degrees of such maps are given by
\begin{equation}\label{eqn:disk M}
	M(y) = \frac{1}{\pi} \oint_{\partial D} y^* d\varphi
\end{equation}
and
\begin{equation}\label{eqn:disk J}
	J^i(y) = -\frac{1}{2\pi i} \oint_{\partial D} y^* dW^i.
\end{equation}
The degrees of $y$ are related to the degrees of $p$ and $q$ by
\begin{equation}\label{eqn:point degrees}
	M(q) - M(p) = M(y), \qquad J^a(q) - J^a(p) = J^a(y)
\end{equation}

Disks only contribute to $Q$ modulo reparametrization along $t$, so let $\# M(p,q,y)$ be the signed count of disks connecting $p$ to $q$ modulo this reparametrization.
\begin{definition} The Floer differential is defined by
\begin{equation}
	Q p = \sum_{\substack{q \in L_0 \cap L_1 \\ M(y)=1, \vec{J}(y)=\vec{0}}} \# M(p,q,y) \, q.
\end{equation}
\end{definition}
From a physics perspective, the intersection points are ground states of the theory and the Floer differential counts instantons tunneling between ground states.

It follows immediately that
\begin{equation}
	Hom_{\mathscr{D}_Y} (L_0, L_1[M]\{\vec{J}\}) = HF^{M,\vec{J}} (L_0, L_1).
\end{equation}
Furthermore, one can use (\ref{eqn:disk M}) and (\ref{eqn:disk J}) to determine the relative degrees of any pair of intersection points regardless of whether $y$ contributes to $Q$. In practice however, there exist simple combinatorial versions of these formulas, derived from the definitions given here, which are far easier to use \cite{Aganagic2}.

\section{Links as A-branes}
\begin{figure}
    \includegraphics{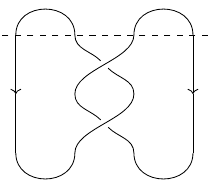}
	\caption{A presentation of the Hopf link as the plat closure a braid with a dashed line dividing it into ``caps" and ``cups".}
	\label{fig:hopf link braid}
\end{figure}

We start with a presentation of an oriented knot or link as the plat closure of a braid $\beta$ on $2d$ strands, such as the Hopf link shown in Figure \ref{fig:hopf link braid}. We then choose a simple Lie algebra $\mathfrak{g}$ and a representation of $\mathfrak{g}$ to color each strand of the knot or link. 

We break the plat closure into two pieces: ``cups" which are segments braided by $\beta$ and ``caps" which are unbraided segments. In Figure \ref{fig:hopf link braid}, the cups and caps are separated by a dashed line. These caps and cups will correspond to special A-branes in a particular A-model. The A-model we want is the one described above, associated to the same Lie algebra $\mathfrak{g}$ and with $d$ marked points on $\mathcal{A}$ colored by representations chosen and $d$ marked points colored by their dual representations. We require $d_a$ to satisfy
\begin{equation}
	\sum_{i=1}^{d} \mu_i = \sum_{a=1}^{rk\mathfrak{g}} d_a e_a
\end{equation}
This data defines a pair $(Y,W)$ and the associated derived Fukaya-Seidel category.

From now on, we will set $\mathfrak{g}=\mathfrak{su}_2$ with each strand colored by the fundamental representation of $\mathfrak{su}_2$. Here, the target space of our theory is $Y = Sym^d{\mathcal{A}}$ and the potential is
\begin{equation}
	W(y) = \lambda_0 \ln f(y) + \lambda_1 \sum_{\alpha=1}^d \ln y_{\alpha}
\end{equation}
with
\begin{equation}
	f(y) = \prod_{\alpha=1}^d \frac{\prod_{i=1}^{2d} (1-a_i/y_{\alpha})}{\prod_{\beta\neq\alpha} (1-y_{\beta}/y_{\alpha})}.
\end{equation}

\begin{figure}
	\centering
	\begin{subfigure}[b]{0.45\textwidth}
		\centering
		\includegraphics{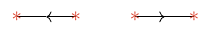}
		\caption{Caps}
		\label{fig:sl2hopflink caps}
	\end{subfigure}
	\begin{subfigure}[b]{0.45\textwidth}
		\centering
		\includegraphics{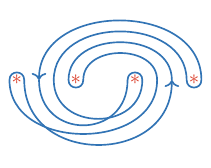}
		\caption{Cups}
		\label{fig:sl2hopflink cups}
	\end{subfigure}
	\caption{Lagrangians corresponding to the caps and cups for the presentation of the Hopf link shown in Figure \ref{fig:hopf link braid} colored by the fundamental representation of $\mathfrak{su_2}$ with $n=2$}
	\label{fig:sl2hopflink caps and cups}
\end{figure}

The caps are then associated to straight line Lagrangians connecting pairs of punctures, such as those shown in Figure \ref{fig:sl2hopflink caps}. To obtain the cups, start with $d$ figure-eight-shaped Lagrangians $E_i$ enclosing pairs of adjacent punctures. We can use $\beta$ to define a map $\mathscr{B}$ from $Y$ to itself by braiding the punctures according to $\beta$. Denote the image of $E_i$ under this map by $\mathscr{B} E_i$. The braided cups are associated to $\mathscr{B}E_\mathcal{U} = \mathscr{B}E_1 \times \ldots \times \mathscr{B}E_n$, the product of the braided figure-eight Lagrangians. An example of $\mathscr{B}E_\mathcal{U}$ for the Hopf link is shown in Figure \ref{fig:sl2hopflink cups}.

\subsection{Link invariants}

Denote the caps by $I_\mathcal{U} = I_1 \times \ldots \times I_n$.

\begin{theorem}\label{thm:invariance}
	The homology groups
	\begin{equation}\label{eqn:invariants}
		Hom^{*,*}_{\mathscr{D}_Y} (\mathscr{B}E_\mathcal{U}, I_\mathcal{U}) = \bigoplus_{M\in\mathbb{Z},\vec{J}\in\mathbb{Z}^2} Hom_{\mathscr{D}_Y} (\mathscr{B}E_\mathcal{U}, I_\mathcal{U}[M]\{\vec{J}\})
	\end{equation}
	are link invariants.
\end{theorem}

We will prove this theorem in Section \ref{sec:algebraic approach} using techniques developed in that section.

\begin{theorem}\label{thm:khovanov}
	The homology groups
	\begin{equation*}
		Hom_{\mathscr{D}_Y} (\mathscr{B}E_\mathcal{U}, I_\mathcal{U}[M]\{J^0\})
	\end{equation*}
	of a link $L$ coincide with the Khovanov homology groups $Kh^{i,j}(L)$ of the same link with the gradings related by
	\begin{equation}
		i = M + 2J^0 + i_0, \qquad j = 2J^0 + j_0
	\end{equation}
	with
	\begin{equation}
		i_0 = \frac{w}{2} - \frac{e}{2}, \qquad j_0 = d + \frac{3w}{2} - \frac{e}{2}
	\end{equation}
	where $d$ is half the number of strands of the braid, $w$ is the writhe the braid, and $e$ is the sum of the exponents of the braid group generators for the presentation chosen.
\end{theorem}
The two theorems are due to Aganagic \cite{Aganagic2} with the general statement of $i_0$ and $j_0$ given in \cite{ALR}.

\begin{figure}
	\includegraphics{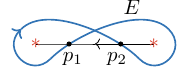}
	\caption{A-branes representing an unknot}
	\label{fig:su2 unknot}
\end{figure}

\begin{example}
Consider the simplest presentation of the unknot, shown in Figure \ref{fig:su2 unknot}, which has $d=1$ and corresponds to the identity braid. There are two intersection points $p_1, p_2 \in E \cap I$. Let $y$ be either of the two disks connecting $p_1$ to $p_2$. Using (\ref{eqn:point degrees}), we find
\begin{equation}
	M(p_2) - M(p_1) = 2
\end{equation}
and
\begin{equation}
	J(p_2) - J(p_1) = -1.
\end{equation}
The absolute degrees are chosen such that $p_1$ has $(M,J^0)=(0,0)$. Once this choice is made for the unknot, the absolute degrees for all other knots are fixed as explained in \cite{ALR}.

Because $p_1$ and $p_2$ do not have the correct degrees to be related by $Q$, we know that $Qp_1=0=Qp_2$. Hence, we find
\begin{equation}
	Hom^{M,J^0}_{\mathscr{D}_Y} (E, I) = \begin{cases}
		\mathbb{Z} & if \ (M,J^0)=(0,0) \ or \ (M,J^0)=(2,-1), \\
		0 & otherwise.
	\end{cases}
\end{equation}
To relate this result to Khovanov homology, note that $w=0$ and $e=0$ so that $i_0=0$ and $j_0=1$. Using Theorem \ref{thm:khovanov}, we find
\begin{equation}
	Kh^{i,j}(L) = \begin{cases}
		\mathbb{Z} & if \ (i,j)=(0,1) \ or \ (i,j)=(0,-1), \\
		0 & otherwise,
	\end{cases}
\end{equation}
which agrees with the known Khovanov homology of the unknot.
\end{example}

\begin{figure}
	\includegraphics{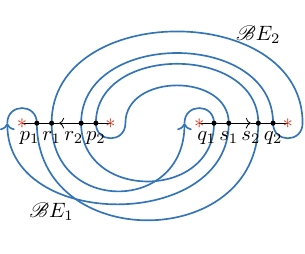}
	\caption{A-branes representing a Hopf link and corresponding to the braid presentation shown in Figure \ref{fig:hopf link braid}}
	\label{fig:su2 hopf link}
\end{figure}

\begin{example}
As a second example, consider the Hopf link shown in Figure \ref{fig:su2 hopf link}, which corresponds to the braid presentation shown in Figure \ref{fig:hopf link braid}. There are eight points in $\mathscr{B}E_\mathcal{U} \cap I_\mathcal{U}$. These points and their degrees are shown in Table \ref{tab:hopf link points}.

\begin{table}[ht]
	\caption{The Maslov degrees $M$ and the equivariant degrees $J^0$ of the intersection points of the branes shown in Figure \ref{fig:su2 hopf link}}
	\label{tab:hopf link points}
	\renewcommand\arraystretch{1.5}
	\noindent\[
	\begin{array}{|c|c|c|c|c|c|c|c|c|}
		\hline
		& p_1 q_1 & p_1 q_2 & p_2 q_1 & p_2 q_2 & r_1 s_1 & r_1 s_2 & r_2 s_1 & r_2 s_2 \\
		\hline
		M & 2 & 4 & 0 & 2 & 0 & 1 & 1 & 2 \\
		\hline
		J^0 & 0 & -2 & 0 & 0 & 1 & 0 & 0 & -1 \\
		\hline
	\end{array}
	\]
\end{table}

One finds the following action of the differential:
\begin{align*}
	Q p_2 q_1 &= r_1 s_2 + r_2 s_1, \\
	Q r_1 s_2 &= -p_1 q_1 - p_2 q_2, \\
	Q r_2 s_1 &= p_1 q_1 + p_2 q_2,
\end{align*}
giving
\begin{equation}
	Hom^{M,J^0}_{\mathscr{D}_Y} (E, I) = \begin{cases}
		\mathbb{Z} & if \ (M,J^0)\in\{(4,-2), (2,-1), (2,0), (0,1)\}, \\
		0 & otherwise.
	\end{cases}
\end{equation}

Looking at Figure \ref{fig:hopf link braid}, we see that this presentation has $w=2$ and $e=2$, so the necessary degree shifts to relate our result to Khovanov homology are $i_0=0$ and $j_0=4$. With these shifts, we recover
\begin{equation}
	Kh^{i,j}(L) = \begin{cases}
		\mathbb{Z} & if \ (i,j)\in\{(0,0), (0,2), (2,4), (2,6)\}, \\
		0 & otherwise,
	\end{cases}
\end{equation}
which is the known Khovanov homology of the positively oriented Hopf link.
\end{example}

\section{An algebraic approach}\label{sec:algebraic approach}
In theory, one could always compute the homology groups $Hom^{*,*}_{\mathscr{D}_Y} (\mathscr{B}E_\mathcal{U}, I_\mathcal{U})$ directly by finding the intersection points of $\mathscr{B}E_\mathcal{U}$ with $I_\mathcal{U}[M]\{\vec{J}\}$ and counting holomorphic disks between intersection points. However, there is no general method for counting these holomorphic disks. We explain an alternative method of computing $Hom^{*,*}_{\mathscr{D}_Y} (\mathscr{B}E_\mathcal{U}, I_\mathcal{U})$, based in algebra rather than geometry, that is more suitable for calculation.

We will use the fact that $\mathscr{D}_Y$ is generated by Lefshetz thimbles to resolve $\mathscr{B}E_\mathcal{U}$ in terms of left thimbles. It will turn out that $I_\mathcal{U}$ are right thimbles, and so calculating Homs between a resolution of $\mathscr{B}E_\mathcal{U}$ and $I_\mathcal{U}$ is easy. A remarkable result is that there is a method for finding a resolution of any $\mathscr{B}E_\mathcal{U}$, giving us an algorithm $Hom^{*,*}_{\mathscr{D}_Y} (\mathscr{B}E_\mathcal{U}, I_\mathcal{U})$ for any link \cite{ALR}.

\subsection{Lefshetz thimbles}
Let $\mathcal{C}$ be a critical point of $\mathrm{Re} \, W$. Each critical point has an associated left Lefshetz thimble $T_\mathcal{C}$ and an associated right Lefshetz thimble $I_\mathcal{C}$.

\begin{definition}
	The left Lefshetz thimble $T_\mathcal{C}$ is the set of all initial conditions for downward gradients flows of $\mathrm{Re} \, W$ that end on $\mathcal{C}$. The right Lefshetz thimble $I_\mathcal{C}$ is the set of all initial conditions for upward gradients flows of $\mathrm{Re} \, W$ that end on $\mathcal{C}$. 
\end{definition}

\begin{figure}
	\includegraphics{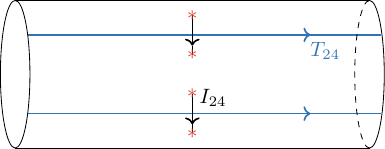}
	\caption{Examples of left and right thimbles in $\mathscr{D}_Y$}
	\label{fig:thimbles}
\end{figure}

It is sufficient for us to know the isotopy classes of the thimbles in $\mathscr{D}_Y$. The left thimbles are isotopic to products of real line Lagrangians running from one end of the cylinder to the other. The right thimbles are isotopic to products of lines running between pairs of punctures. Examples of left and right thimbles are shown in Figure \ref{fig:thimbles}. The left thimbles are numbered such that $T_i$ is a line passing between punctures $a_i$ and $a_{i-1}$, and $T_{i_1\ldots i_d}$ is the product $T_{i_1} \times \ldots \times T_{i_d}$. The right thimbles are numbered similarly.

Left and right Lefshetz thimbles are dual in the sense that they only intersect if they are associated to the same critical point. Hence, we have
\begin{equation}
	Hom_{\mathscr{D}_Y} (T_\mathcal{C}, I_{\mathcal{C}'}) = \delta_{\mathcal{C}, \mathcal{C}'}.
\end{equation}

\begin{figure}
	\includegraphics{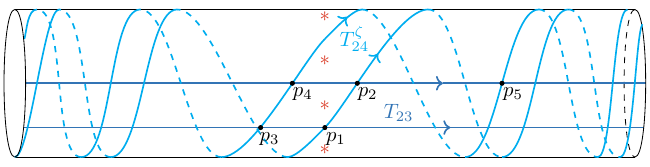}
	\caption{An example of a thimble $T_{23}$ and a wrapped thimble $T_{24}^\zeta$ whose intersection points give $Hom_{\mathscr{D}_Y}(T_{24}, T_{23})$}
	\label{fig:wrapped thimbles}
\end{figure}

To compute link invariants, it will be necessary to compute Homs between the right thimbles as well. Because these thimbles are noncompact, we define Homs between them using wrapping. Let $T_\mathcal{C}^\zeta$ be the right thimble obtained by replacing $W$ with $W e^{-i\zeta}$ \cite{HIV}. The effect of this replacement is shown in Figure \ref{fig:wrapped thimbles}. We then define morphisms between right thimbles as
\begin{equation}
	Hom_{\mathscr{D}_Y} (T_\mathcal{C}, T_{\mathcal{C}'}) = HF^{0,0} (T_\mathcal{C}^\zeta, T_{\mathcal{C}'}).
\end{equation}
Morphisms between left thimbles are defined analogously.

\begin{figure}
	\includegraphics{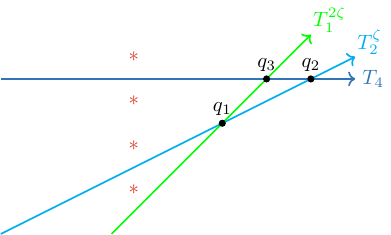}
	\caption{A holomorphic triangle giving the relation $q_2 \cdot q_1 = q_3$}
	\label{fig:holomorphic triangle}
\end{figure}

Morphisms compose via
\begin{equation}
	Hom_{\mathscr{D}_Y}(T_j, T_k) \otimes Hom_{\mathscr{D}_Y}(T_i, T_j) \to Hom_{\mathscr{D}_Y}(T_i, T_k),
\end{equation}
which, in terms of Floer theory, is
\begin{equation}
	HF^{0,0}(T_j^\zeta, T_k) \otimes HF^{0,0}(T_i^{2\zeta}, T_j^\zeta) \to HF^{0,0}(T_i^{2\zeta}, T_k).
\end{equation}
Composition of morphisms amounts to counting holomorphic triangles such as the one shown in Figure \ref{fig:holomorphic triangle}.

Morphisms between all left thimbles form a graded, associative algebra $A$. It is best described as a strand algebra on a cylinder with $2d$ red strands corresponding to the fixed positions of the punctures and $d$ blue strands corresponding to the positions of the thimbles. The thimble $T_i$ is represented by a blue strand between the $i$th and $(i-1)$th red strands. The thimble $T_{i_1 \ldots i_d}$ is represented by the $d$ blue strands representing $T_{i_1},\ldots,T_{i_d}$.

Consider a morphism from $T_{\mathcal{C}}$ to $T_{\mathcal{C}'}$ and its associated intersection point. The ordering of the strands at the base of the strand algebra cylinder corresponds to $T_{\mathcal{C}}$ and the ordering of the strands at the top of the strand algebra cylinder corresponds to $T_{\mathcal{C}'}$. The intersection point uniquely specifies how the strands connect the base of the cylinder to the top of the cylinder.

For example, the three intersection points $p_1 p_4, p_2 p_3, p_1 p_5 \in Hom_{\mathscr{D}_Y}(T_{24}, T_{23})$ shown in Figure \ref{fig:wrapped thimbles} correspond to the diagrams
\begin{equation*}
	\vcenter{\hbox{\includegraphics{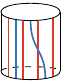}}} \qquad \qquad \vcenter{\hbox{\includegraphics{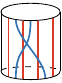}}} \qquad \qquad \vcenter{\hbox{\includegraphics{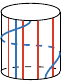}}}
\end{equation*}
respectively.

Composition of morphisms is given by stacking the corresponding cylinders and rescaling. For example, the relation $q_2 \cdot q_1 = q_3$ shown in Figure \ref{fig:holomorphic triangle} is given in terms of strand diagrams as
\begin{equation}
    \vcenter{\hbox{\includegraphics{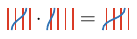}}}
\end{equation}
where we have omitted drawing the cylinders for convenience. For knots in $\mathbb{R}^3$, one will never encounter algebra elements that wrap behind the cylinder.

Altogether, the strand algebra is generated by certain ``bits":
\begin{equation*}
	\vcenter{\hbox{\includegraphics{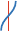}}} \qquad \qquad \vcenter{\hbox{\includegraphics{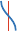}}} \qquad \qquad \vcenter{\hbox{\includegraphics{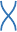}}}
\end{equation*}
that have $J^0=0$, $J^0=1$, and $J^0=-1$, respectively. The equivariant degree of each strand algebra element is the sum of the equivariant degrees of the ``bits" that compose it.

It turns out that the branes in $\mathscr{D}_Y$ have slightly more structure than we have explained so far. Each brane comes equipped with a local system of modules of a graded algebra $\mathcal{B}$. A morphism between branes will also involve homomorphisms of their local systems at the corresponding intersection point. The local systems can be viewed as coming from a larger A-model $(\mathcal{Y}, \mathcal{W})$ where $\mathcal{Y}$ fibers over $Y$ with $(\mathbb{C}^\times)^d$ fibers. Each thimble $T_\mathcal{C}$ corresponds to a thimble $\mathcal{T}_\mathcal{C}$ in $\mathcal{Y}$ where $T_\mathcal{C}$ and $\mathcal{T}_\mathcal{C}$ agree on $Y$. From this perspective, one can view homomorphisms of $\mathcal{B}$-modules as coming from the intersections of thimbles in the fibers over their intersection points in $Y$. For details, see \cite{ADLZ}.

For our purposes, it is sufficient to only consider the effect of the local systems on the thimble algebra. The effect is to add another generator, a dot with $J^0=1$ that may be placed on any blue strand.

The strand algebra contains several nontrivial relations that are proven in \cite{ADLZ} by counting holomorphic disks in the derived Fukaya-Seidel category associated to $(\mathcal{Y}, \mathcal{W})$. The relations are
\begin{align*}
	&1.\; \vcenter{\hbox{\includegraphics{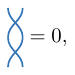}}} &&	2.\; \vcenter{\hbox{\includegraphics{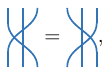}}} \\
	&3.\; \vcenter{\hbox{\includegraphics{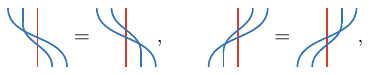}}} \\
	&4.\; \vcenter{\hbox{\includegraphics{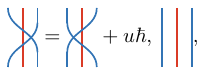}}} && 5.\; \vcenter{\hbox{\includegraphics{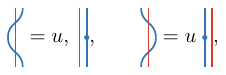}}} \\
	&6.\; \vcenter{\hbox{\includegraphics{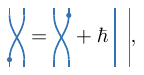}}} && 7.\; \vcenter{\hbox{\includegraphics{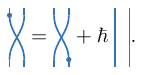}}}
\end{align*}
One may always rescale the algebra generators to set $u=1$ and $\hbar=1$, but we leave them in for convenience. With $u=1$, these relations are the same as those for the KLRW algebra \cite{Webster}. The fact that the thimble algebra coincides with the KLRW algebra is expected from homological mirror symmetry \cite{Aganagic2}.

\begin{remark}
	Let $\Delta$ be the divisor of $\prod_{\alpha < \beta} (y_\alpha - y_\beta)$. The algebra at $\hbar=0$ is the strand algebra on the space $Y_0 = Y \backslash \Delta$, the complement of the diagonal $\Delta$ in $Y$. The theory on $Y_0$ is similar to the theory on $Y$ except that disks cannot pass through any point with $y_\alpha = y_\beta$, so every disk is a product of one-dimensional disks.
\end{remark}

\begin{remark}
	Let $D_a$ be the divisor of $\prod_{\alpha,i} (y_\alpha - a_i)$. The algebra at $u=0$ and $\hbar=0$ is the strand algebra on the space $Y_0 \backslash D_a$. In this theory, disks are not allowed to pass through the punctures. It corresponds to all branes being equipped with a trivial local system.
\end{remark}

\subsection{Resolutions of A-branes}
Every A-brane in $\mathscr{D}_Y$ admits a resolution in terms of thimbles. Resolutions of one-dimensional branes can be easily written down from geometry. We will use these one-dimensional resolutions as building blocks to find resolutions of higher dimensional branes.

\subsubsection{Resolutions of one-dimensional Lagrangians}
\begin{figure}
	\centering
	\includegraphics{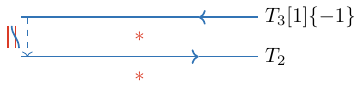} \qquad \qquad
	\includegraphics{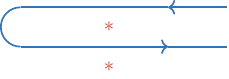}
	\caption{On the left, two thimbles with a map between them and on the right, the Lagrangian corresponding to the cone over that map}
	\label{fig:cone}
\end{figure}

We use mapping cones to construct of one-dimensional Lagrangians.
\begin{definition}
	Let $(L_1, \delta^{L_1})$ and $(L_2, \delta^{L_2})$ be chain complexes and let $f:L_1 \to L_2$ be a chain map. The cone over $f$ is the complex $L_1[1] \oplus L_2$ equipped with the differential
	\begin{equation}
		\delta = \begin{pmatrix} \delta^{L_1} & 0 \\ f & -\delta^{L_2} \end{pmatrix}
	\end{equation}
\end{definition}
In one dimension, the cone over $f:L_1 \to L_2$ amount to gluing $L_1[1]$ and $L_2$ at the point $p \in Hom_{\mathscr{D}_Y}(L_1,L_2)$. For example, consider the map $\vcenter{\hbox{\includegraphics{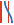}}}:T_3\{-1\} \to T_2$. The cone over $\vcenter{\hbox{\includegraphics{b23}}}$ is 
\begin{equation}
	T_3\{-1\} \xrightarrow{\vcenter{\hbox{\includegraphics{b23}}}} T_2.
\end{equation}
Geometrically, this cone is the resolution of the brane shown in Figure \ref{fig:cone}.

In $\mathscr{D}_{Y_0 \backslash D_a}$, resolutions of more complicated one-dimensional branes can be built by gluing thimbles along maps. For resolutions of branes in $\mathscr{D}_{Y_0}$ however, we need to add additional maps. To find a resolution of a brane in $\mathscr{D}_{Y_0}$, we start with a resolution of a brane in $\mathscr{D}_{Y_0 \backslash D_a}$. Call the differential $\delta$, and note that $\delta^2 \vert_{u=0} = 0$. We then make an ansatz for the differential in $\mathscr{D}_{Y_0}$ by adding to $\delta$ all possible maps with dots as allowed by degree with arbitrary coefficients. We then use the condition $\delta^2 = 0$ to fix the coefficients. This uniquely determines a brane in $\mathscr{D}_{Y_0}$. Since the brane is one-dimensional, it is also a brane in $\mathscr{D}_{Y}$.

\begin{example}
Consider the figure-eight brane $E$ shown in Figure \ref{fig:su2 unknot}, which we refer to as $E_2$ from now on. We can break $E_2$ into thimbles and maps between them as shown in Figure \ref{fig:cup resolution}. Assembling the thimbles and maps into a complex, we find
\begin{equation}
	E_2 \qquad \cong \qquad
	T_2\{-1\}
	\xrightarrow{\vcenter{\hbox{\includegraphics{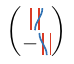}}}}
	\begin{matrix} T_3\{-1\} \\ \oplus \\ T_1 \end{matrix}
	\xrightarrow{\vcenter{\hbox{\includegraphics{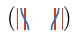}}}}
	T_2.
\end{equation}
This differential squares to zero in $\mathscr{D}_Y$, so we are done.

\begin{figure}
    \includegraphics{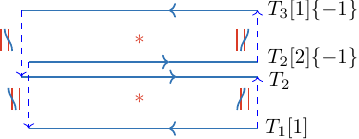}
	\caption{The thimbles and maps between them making up the resolution of the brane $E$}
	\label{fig:cup resolution}
\end{figure}

As a check, note that applying $Hom^{*,*}_{\mathscr{D}_Y} (-, I_2)$ to $E_2$ recovers the correct Khovanov homology of the unknot. The resolution of $E_2$ contains two copies of $T_2$, so we find
\begin{equation}
	Hom^{M,J^0}_{\mathscr{D}_Y} (E_2, I_2) = \begin{cases}
		\mathbb{Z} & if \ (M,J^0)=(0,0) \ or \ (M,J^0)=(2,-1), \\
		0 & otherwise,
	\end{cases}
\end{equation}
agreeing with what we found previously.
\end{example}

\begin{figure}
	\includegraphics{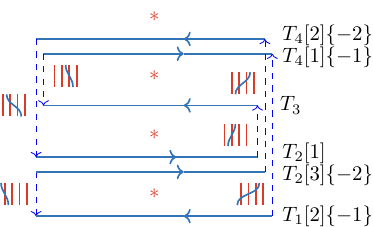}
	\caption{The thimbles and maps between them making up the resolution of the brane $\mathscr{B}E_1$ shown in Figure \ref{fig:su2 hopf link}}
	\label{fig:hopf link resolution 1}
\end{figure}

\begin{example}
	Consider the brane $\mathscr{B}E_1$ shown in Figure \ref{fig:su2 hopf link}. We can decompose $\mathscr{B}E_1$ as thimbles and maps between them as shown in Figure \ref{fig:hopf link resolution 1}. Assembling these thimbles and maps into a chain complex gives
    \begin{equation}
		T_2\{-2\}
		\xrightarrow{\vcenter{\hbox{\includegraphics{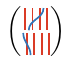}}}}
		\begin{matrix} T_4\{-2\} \\ \oplus \\ T_1\{-1\} \end{matrix}
		\xrightarrow{\vcenter{\hbox{\includegraphics{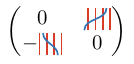}}}}
		\begin{matrix} T_4\{-1\} \\ \oplus \\ T_2 \end{matrix}
		\xrightarrow{\vcenter{\hbox{\includegraphics{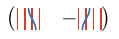}}}}
		T_3.
	\end{equation}
	This differential squares to zero when $u=0$ but not when $u \neq 0$, so this complex is a resolution of the brane in $\mathscr{D}_{Y_0 \backslash D_a}$ but not in $\mathscr{D}_{Y}$. In order to find a resolution in $\mathscr{D}_{Y}$, we must add dotted correction terms. Adding in all possible dotted corrections with undetermined coefficients $x_i$, the complex becomes
    \begin{equation}
		T_2\{-2\}
		\xrightarrow{\vcenter{\hbox{\includegraphics{BE1d3.pdf}}}}
		\begin{matrix} T_4\{-2\} \\ \oplus \\ T_1\{-1\} \end{matrix}
		\xrightarrow{\vcenter{\hbox{\includegraphics{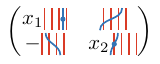}}}}
		\begin{matrix} T_4\{-1\} \\ \oplus \\ T_2 \end{matrix}
		\xrightarrow{\vcenter{\hbox{\includegraphics{BE1d1.pdf}}}}
		T_3.
	\end{equation}
	Requiring $\delta^2=0$ fixes $x_1=-u$ and $x_2=u$, giving us a resolution of $\mathscr{B}E_1$ in $\mathscr{D}_{Y}$. One can follow a similar procedure to resolve $\mathscr{B}E_2$.
\end{example}

\subsubsection{Resolutions of higher dimensional Lagrangians}
Given a $d$-dimensional Lagrangian $L$, we first break it into a product of one-dimensional Lagrangians $L_i$ for $i=1,\ldots,d$. We then find a resolution of each one-dimensional Lagrangian using the methods just described. To construct a resolution $L$, we iteratively take the product of $L_1 \times \ldots \times L_{i}$ with $L_{i+1}$, adding in extra maps to ensure $\delta^2=0$ at each step. We will briefly outline how this construction goes. More details can be found in \cite{ALR}.

Let $\delta^{L_i}$ be the differential on $L_i$. Start by considering the product $L_1 \times L_2$ where differential acts by
\begin{equation}
	\delta_\mathrm{geo}^{L_1 \times L_2}(\ell_1, \ell_2) = (\delta^{L_1} \ell_1, \ell_2) + (-1)^{M(\ell_1)} (\ell_1, \delta^{L_2} \ell_2).
\end{equation}
This differential does not necessarily satisfy $\left( \delta_\mathrm{geo}^{L_1 \times L_2} \right)^2 = 0$, so we add a correction term $\tilde\delta^{L_1 \times L_2}$ to get a differential
\begin{equation}
	\delta^{L_1 \times L_2} = \delta_\mathrm{geo}^{L_1 \times L_2} + \tilde\delta^{L_1 \times L_2}
\end{equation}
that does square to zero. Repeat this construction using $L_1 \times L_2$ and $L_3$ to construct $L_1 \times L_2 \times L_3$ and so on until arriving at $L_1 \times \ldots \times L_d$ equipped with a differential $\delta^{L_1 \times \ldots \times L_d}$ satisfying $\left( \delta^{L_1 \times \ldots \times L_d} \right)^2=0$.

To find $\tilde\delta$ at each step, note that the possible maps it could contain are restricted by the equivariant degrees of the thimbles in $L_1 \times \ldots \times L_d$. As a result, there are finite numbers of possibilities for $\tilde\delta$. One can make an ansatz for $\tilde\delta$ by adding all possible maps with undetermined coefficients and then using the condition $\delta^2 = 0$ to fix the coefficients. We only add maps where there is not already a map in $\delta_\mathrm{geo}$. 

\begin{example}
	Consider the two-dimensional brane $L_1 \times L_2$ shown in Figure \ref{fig:d2 example}. The resolutions of the one-dimensional branes $L_1 \cong E_2$ and $L_2$ are easily found to be
    \begin{equation}
		L_1 \qquad \cong \qquad
		T_2\{-1\}
		\xrightarrow{\vcenter{\hbox{\includegraphics{E2d2.pdf}}}}
		\begin{matrix} T_3\{-1\} \\ \oplus \\ T_1 \end{matrix}
		\xrightarrow{\vcenter{\hbox{\includegraphics{E2d1.pdf}}}}
		T_2,
	\end{equation}
	and
    \begin{equation}
		L_2 \qquad \cong \qquad T_3\{-2\} \xrightarrow{\vcenter{\hbox{\includegraphics{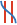}}}} T_1.
	\end{equation}
	Taking the product of $L_1$ with $L_2$, we find
    \begin{equation}
		T_{23}\{-2\}
		\xrightarrow{\vcenter{\hbox{\includegraphics{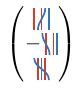}}}}
		\begin{matrix} T_{33}\{-2\} \\ \oplus \\ T_{13}\{-1\} \\ \oplus \\ T_{12}\{-1\} \end{matrix}
		\xrightarrow{\vcenter{\hbox{\includegraphics{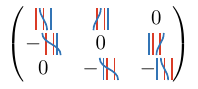}}}}
		\begin{matrix} T_{23}\{-1\} \\ \oplus \\T_{13}\{-1\} \\ \oplus \\ T_{11} \end{matrix}
		\xrightarrow{\vcenter{\hbox{\includegraphics{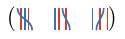}}}}
		T_{12}.
	\end{equation}
	Notice that the equivariant degrees are different from the naive expectation because the crossing of two thimbles in the strand algebra has nonzero equivariant degree. The equivariant degrees of the thimbles are adjusted to stay consistent with the equivariant degrees of the maps.
	
	\begin{figure}
		\centering
		\begin{subfigure}[b]{0.45\textwidth}
			\centering
			\includegraphics{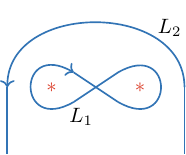}
			\caption{Simple two-dimensional brane}
			\label{fig:d2 example}
		\end{subfigure}
		\begin{subfigure}[b]{0.45\textwidth}
			\centering
			\includegraphics{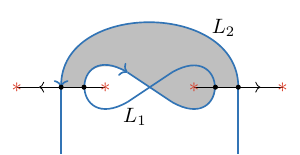}
			\caption{Disk from intersection with $I_{13}$}
			\label{fig:d2 example hom}
		\end{subfigure}
		\caption{A simple example of a two-dimensional brane $L_1 \times L_2$ and the holomorphic disk that contributes to the Floer differential on intersection of $L_1 \times L_2$ with $I_{13}$}
		\label{fig:d2 example with hom}
	\end{figure}
	
	One can verify that the differential on the product complex squares to zero only in $\mathscr{D}_{Y_0}$ but not in $\mathscr{D}_Y$, so we must add a correction to the differential. There are only two possible corrections allowed by degrees. With undetermined coefficients $x_i$ for the correction terms, the differential is
    \begin{equation}
		T_{23}\{-2\}
		\xrightarrow{\vcenter{\hbox{\includegraphics{L12d3.pdf}}}}
		\begin{matrix} T_{33}\{-2\} \\ \oplus \\ T_{13}\{-1\} \\ \oplus \\ T_{12}\{-1\} \end{matrix}
		\xrightarrow{\vcenter{\hbox{\includegraphics{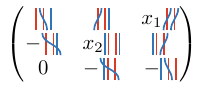}}}}
		\begin{matrix} T_{23}\{-1\} \\ \oplus \\T_{13}\{-1\} \\ \oplus \\ T_{11} \end{matrix}
		\xrightarrow{\vcenter{\hbox{\includegraphics{L12d1.pdf}}}}
		T_{12}.
	\end{equation}
	Solving $\delta^2=0$ in $\mathscr{D}_Y$ fixes $x_1=0$ and $x_2=-u\hbar$, giving us a resolution for $L_1 \times L_2$.
	
	As a check, we apply $Hom^{*,*}_{\mathscr{D}_Y} (-, I_{13})$ to the resolution and compare it to the geometric result. We find
	\begin{equation}
		Hom^{*,*}_{\mathscr{D}_Y} (L_1 \times L_2, I_{13}) = H^*\left(\mathbb{Z} \xrightarrow{u\hbar} \mathbb{Z}\right) = 0.
	\end{equation}
	The corresponding geometric brane configuration is shown in Figure \ref{fig:d2 example hom}. There are two intersection points connected by a holomorphic disk, shaded gray, so the resulting Floer complex matches the result we obtained algebraically.
\end{example}

\begin{remark}
	One may have noticed that the resolution of $L_1 \times L_2$ may have just as easily been computed by finding a chain map $f: E_2 \times T_3 \{-1\} \to T_1 \times E_2$ and taking a cone over $f$. In most examples however, the method we used here is more tractable, so we felt there was value in explaining how it works for a simple example.
\end{remark}

For more complicated examples, it is easiest to work iteratively in $\hbar$ to solve $\delta^2 = 0$. Expand $\delta$ as
\begin{equation}
	\delta = \sum_{k=0}^\infty \hbar^k \delta_k
\end{equation}
so that
\begin{equation}
	\delta^2 = \delta_0^2 + \hbar (\delta_0 \delta_1 + \delta_1 \delta_0) + \hbar^2 (\delta_1^2 + \delta_0 \delta_2 + \delta_2 \delta_0) + \ldots = 0
\end{equation}
is satisfied at each order in $\hbar$. Note that $\delta_\mathrm{geo}^2\vert_{\hbar=0} = 0$, so we can identify $\delta_0 = \delta_\mathrm{geo}$. Then $\tilde\delta$ contains terms of at least order $\hbar$ and we identify
\begin{equation}
	\tilde\delta = \sum_{k=1}^\infty \hbar^k \delta_k.
\end{equation}
Also notice that at order $k$, the equation $\delta^2=0$ is linear in $\delta_k$ once $\delta_0,\ldots,\delta_{k-1}$ are fixed, so finding the solution very tractable on a computer.

Addressing the issue of uniqueness of this construction requires considerable care. We claim that this construction is unique and refer the reader to \cite{ALR} for the proof.

\subsection{Proof of invariance}\label{ssec:invariance}
We give a brief sketch of a proof of Theorem \ref{thm:invariance} here. For a more detailed version, see \cite{ALR}, or for a proof based on homological mirror symmetry, see \cite{Aganagic2}.

\begin{figure}
    \includegraphics{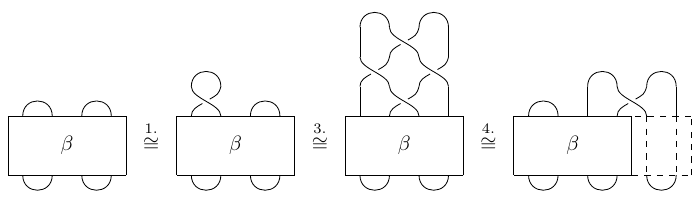} \\
    \includegraphics{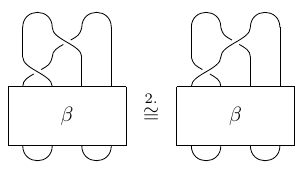}
	\caption{The four relations that must be satisfied for a quantity to be a link invariant}
	\label{fig:invariance moves}
\end{figure}

\begin{figure}
    2. $\vcenter{\hbox{\includegraphics{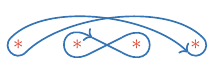}}} \enspace \cong \enspace \vcenter{\hbox{\includegraphics{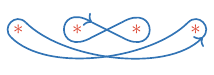}}}$ \\
    4. $\vcenter{\hbox{\includegraphics{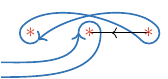}}} \enspace \cong \enspace \vcenter{\hbox{\includegraphics{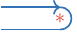}}}$
	\caption{Invariance relations 2 and 4 shown in terms of branes}
	\label{fig:brane moves}
\end{figure}

\begin{figure}
    $\vcenter{\hbox{\includegraphics{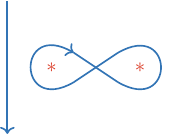}}} \enspace \cong \enspace \vcenter{\hbox{\includegraphics{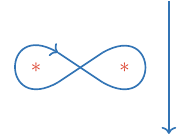}}} \enspace \{-1\}$
	\caption{Two equivalent branes in $\mathscr{D}_Y$}
	\label{fig:slide move}
\end{figure}

\begin{proof}[Sketch of proof of Theorem \ref{thm:invariance}]
	We will show that $Hom^{*,*}_{\mathscr{D}_Y} (\mathscr{B}E_\mathcal{U}, I_\mathcal{U})$ satisfy the four relations shown in Figure \ref{fig:invariance moves}, which is sufficient to prove invariance \cite{Bigelow}, \cite{Birman}. The fact that relations 1 and 3 are satisfied is immediate. One only has to be careful to keep track of the overall degree shifts.
	
	Relations 2 and 4 require more work. These relations are shown in terms of A-branes in Figure \ref{fig:brane moves}. Both of these relations follow from the relation shown in Figure \ref{fig:slide move}. Hence, it is sufficient to prove this relation. One can do so by constructing resolutions of the two branes using the method described above and then explicitly showing that the resolutions are homotopy equivalent. 
\end{proof}

\bibliographystyle{amsalpha}

\end{document}